\documentclass{svjour3}                     
\smartqed  
\usepackage{graphicx}
\usepackage{natbib}
\usepackage{hyperref}
\usepackage{longtable}
\usepackage{amsmath}
\providecommand{\class}[1]{{\it #1}}

\journalname{	Experimental Astronomy - DOI: 10.1007/s10686-011-9248-z}
\begin{document}

\date{Received: date / Accepted: date}
\title{Implementing the Gaia Astrometric Global Iterative Solution (AGIS) in Java}

\titlerunning{Implementing AGIS}        

\author{William O'Mullane$^{1}$ and Uwe Lammers$^{1}$ and Lennart Lindegren$^{2}$ 
and Jose Hernandez$^{1}$ and David Hobbs$^{2}$
}

\authorrunning{O'Mullane et al.} 

\institute{
$^1$European Space Astronomy Centre (ESAC),P.O. Box -- Apdo. de correos 78, ES-28691 Villanueva de la Ca{\~n}ada, Madrid\\
\email{William.OMullane, Uwe.Lammers@sciops.esa.int, Jose.Hernandez@sciops.esa.int}\\
$^2$Lund Observatory, Lund University, Box 43, SE-22100 Lund
\email{Lennart.Lindegren, David.Hobbs@astro.lu.se}
}

\date{Aug. 2011}

\maketitle
\abstract{
This paper provides a description of the Java software framework which has been
constructed to run the Astrometric Global Iterative Solution for the Gaia
mission. This is the mathematical framework to provide the rigid reference
frame for Gaia observations from the Gaia data itself. This process makes 
Gaia a self calibrated, and input catalogue independent, mission.
The framework is highly distributed typically running on a cluster of machines
with a database back end. All code is written in the Java language.  
We describe the overall architecture and some of the details of the implementation.
 
}

\section{Introduction  }\label{sect:intro}

Astrometry is one of the oldest pursuits in science. The measurement of positions
and later motions of celestial bodies has been an occupation for millennia. The
most famous, but now lost, star catalogue of the Antiquity was compiled around 129~BC
by Hipparchus \citep{2005JHA....36..167S}, whose name is
echoed in the Hipparcos mission \citep{hip:catalogue} which brought the first space-based 
astrometry. Gaia continues in this ancient tradition using the most modern
of techniques.

ESA is due to launch the $\sim$2000~kg Gaia satellite in 2013 on a Soyuz-Fregat
rocket to the L2 point some 1.5~million km from earth. It consists of an astrometric 
instrument with two viewing directions, complemented by photometric and radial-velocity 
instruments providing astrophysical information and allowing it to build a phase-space 
map of our galaxy.

Over its five-year mission Gaia will obtain astrometric and photometric data for about
a thousand million sources (stars, quasars, and other point-like objects); a subset of 
about 250~million of the brighter sources will also 
be observed spectrographically. Gaia will use a mosaic of CCD detectors operated in a
drift-scanning mode throughout the five years, producing an average of approximately 
700 individual CCD observations of each source and covering the entire sky 
three-fold every six months. For more detailed overviews of the Gaia project and its 
science goals we refer to, e.g., \citet{2005ESASP.576.....T}, \citet{2008IAUS..248..217L}, 
\citet{2008AN....329..875J}, \citet{2010IAUS..261..296L} and \citet{2010SPIE.7731E..35D}.

A central part of the data processing for Gaia is the so-called Astrometric Global Iterative
Solution (AGIS), which transforms the $\sim${}$10^{12}$ individual observations into an 
astrometric catalogue of unprecedented accuracy. The full mathematical details of AGIS 
are given elsewhere \citep{LL08} and are only briefly referred to below. In the present 
paper we discuss the overall architecture of the processing framework that is being set
up to carry out this huge task, as well as some details of the implementation.

When reading this paper it should be borne in mind that the word {\em Object} will
be used in the sense that is normal in computer science or object-oriented programming. 
It should not be confused with an astronomical object, for which, in general, we 
use the term {\em Source}. For improved clarity, names of classes and methods are 
generally set in italics when they appear in regular text.

\section{The Gaia Astrometric Global Iterative Solution (AGIS)}

\subsection{Astrometry as a minimization problem}\label{sec:intromath}

In \citet{hip:catalogue} the general principle of a global astrometric mission is
succinctly formulated as the minimization problem:
\begin{equation}
        \min_{\vec{s},\vec{n}} \| \vec{g}^\text{obs} - \vec{g}^\text{calc}(\vec{s},\vec{n})\|_{M}
        \label{eq:generalobs}
\end{equation} 
where $\vec{g}^\text{obs}$ is the vector of all the observations (measurements),
$\vec{g}^\text{calc}$ the corresponding calculated values, and the norm is calculated in
some metric $M$ that takes into account the different weights of the observations.

The vector $\vec{s}$ represents the (unknown) astrometric parameters of the sources. As described
in detail in \citep{LL08}, each source $i$ is modelled in terms of six astrometric parameters, 
namely:

\begin{tabular}{l p{0.80 \textwidth}}
$\alpha_i$ & right ascension at a given reference time, i.e., the longitude-like 
coordinate along the celestial equator\\
$\delta_i$ &  right ascension at a given reference time, i.e., the angular distance from 
the celestial equator (positive towards north)\\
$\varpi_{i}$ & annual parallax, inversely proportional to distance from the sun\\
$\mu_{\alpha*i}$ & ($=\mu_{\alpha}\cos\delta$) proper motion in right ascension, i.e., 
the annual change in $\alpha$ times $\cos\delta$\\
$\mu_{\delta i}$ & proper motion in declination, i.e., the annual change in $\delta$\\
$v_{r i}$ & radial velocity, i.e., the rate of change of the distance to the source.\\
\end{tabular}

\noindent
The radial velocity $v_{r i}$ is best determined spectroscopically, using the Doppler shift of 
spectral lines, and is not included among the unknowns to be determined by the
astrometric solution. The vector $\vec{s}$ therefore contains 5 unknowns for each
source. The astrometric solution will operate on a subset of about 10\% of the sources
known as the primary sources (see Sect.~\ref{sec:secondary}), so the total number of astrometric
unknowns is some $5\times 10^8$.

The vector $\vec{n}$ contains the nuisance parameters, i.e., all other parameters that 
need to be determined simultaneously with $\vec{s}$, using the same observations,
because they cannot be measured accurately enough by other means. These include
the satellite attitude, the geometric calibration of the instrument, and a few global 
parameters. Their total number is of the order of $10^7$.

\subsection{Iterative solution}\label{sect:iteration}

Equation~(\ref{eq:generalobs}) means that the model, encapsulated by the function 
$\vec{g}^\text{calc}$, is fitted to the observations by adjustment of the parameters 
$\vec{s}$ and $\vec{n}$. To directly
fit all parameters is infeasible, considering their number in excess of $n=5\times 10^8$.
A brute-force direct solution would
require about $n^3/6 \sim 2\times 10^{25}$ FLOPs and the normal equations matrix 
would occupy about $n^2/2 \sim 10^{17}$ doubles or 1~exabyte (1~million TB) of storage. 
Rather than a direct solution we take a block iterative approach.

We model the effects of the source, attitude, calibration and global parameters independently, 
treating the  dependencies as given. Hence to solve for the astrometric parameters 
of a source we
assume some attitude, calibration and global parameters; then  for calibration
we assume the global, attitude and astrometric parameters, and so on. The order in which 
this is done should in principle not matter although solving the astrometry for the
individual sources first is logical and has some advantages (Sect.~\ref{sec:secondary}). 
Hence the solution 
would involve four relatively independent blocks of equations, where each takes
the form of the general minimization problem of Eq.~(\ref{eq:generalobs}), 
although only for a subset of the parameters. The four blocks are referred to as the
Source Update, Attitude Update, Calibration Update, and Global Update.

The convergence properties of this kind of (simple) iterative solution were essentially
unknown when the Gaia data processing system was first planned. Although it
was {\em felt} that it should converge, there was no proof of even that. The early work
outlined in \citet{1999BaltA...8...57O} was a first indication that convergence in a few
tens of iterations should be possible. Subsequent experiments have shown that the
iterations do indeed converge, although slowly, and that the convergence speed can
be improved considerably by modifying the updates to take into account previous
updates. The current solution method, based on the conjugate gradients algorithm, 
converges and effectively removes all systematic errors in the initial catalogue data 
in some 40--100 iterations, when applied to simulated data \citep{bombrun+09}. 
In practice one must iterate until the updates become very small, and further work 
continues to define an exact convergence criterion.

The efficient software implementation of the block iterative solution
is challenging. A first attempt for such a solution during the Hipparcos
data processing was abandoned. A  basic proof of concept, actually more a pseudo
implementation,  using again Hipparcos data and a database management system, was
presented in  \citet{1999BaltA...8...57O}. A good deal of
effort went into scaling this up to Gaia dimensions until finally a degree of
success was gained by the ESAC group \citep{2006astro.ph.11885O} in 2005. It is
this ESAC framework which is presented here and which shall continue to be 
developed up to and even after the launch of Gaia.

\section{Overview of the AGIS data processing system} \label{sect:agisoverview}

\begin{figure}
\begin{center}
\includegraphics[scale=0.47,trim=0 22cm 0 0]{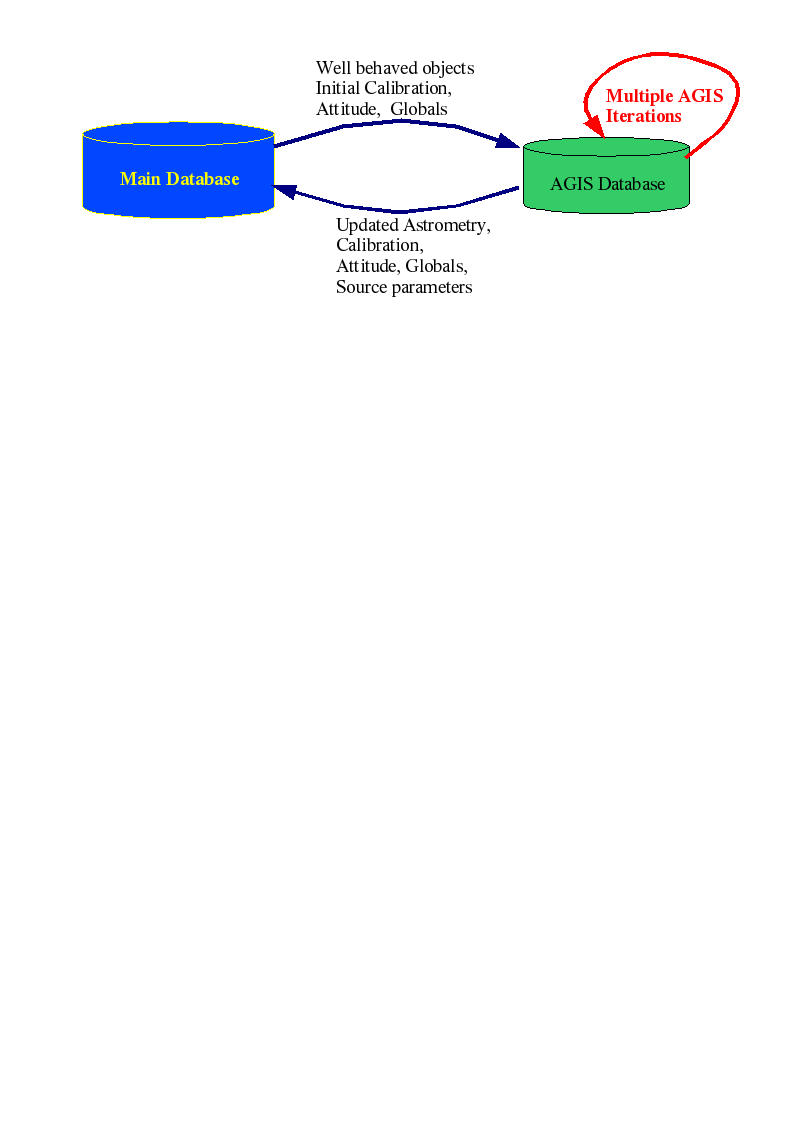}
\caption{ \label{fig:AGIScontext} AGIS, like other Gaia processing systems,
extracts data from the Main Database. Updated results are fed back to the Main
Database and merged with results coming from other processing systems.}
\end{center}
\end{figure}

AGIS is just one of many parts of the Gaia processing,
a central or core part certainly but still a part. In the overall design of the
Gaia processing system the Main Database is the central repository of all information.
Figure~\ref{fig:AGIScontext} depicts AGIS
is this broader context with the Main Database.

\begin{figure}
\begin{center}
\includegraphics[scale=0.35]{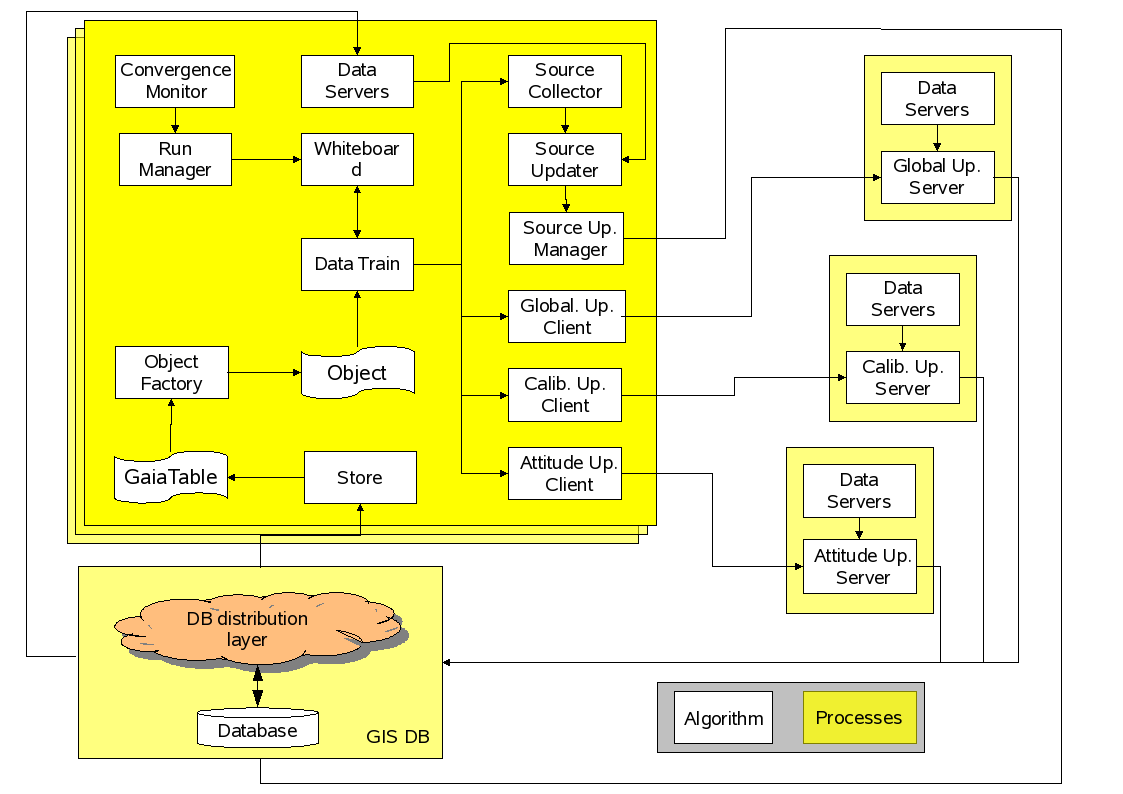}
\caption{ \label{fig:agis} 
Logical overview of AGIS. The many processes of AGIS run on many 
different machines (not shown here).
The large box on the left represents the {\it DataTrain}, of which there may be a great
number running. On the right are the update servers, of which there may be only
one of each kind running in the entire system. A database management system underpins all of these
processes.}
\end{center}
\end{figure}

A simplified overall AGIS picture is presented in Fig.~\ref{fig:agis}. Each of
the components in the picture may run on practically any regular machine apart from the Attitude
Update  Server, which requires a little more memory (of the order of 16~GB). The {\it DataTrain}, as mediator, is
seen in the middle of the left box and is explained in some detail in
Sect.~\ref{sect:datatrain}. 
The database systems -- currently InterSystems Cach{\'e}, Oracle Real
Application Clusters, or (for small data sets) Apache Derby -- may also run on several machines (or nodes)  to
improve data access performance. The data access and storage is abstracted
through the {\it Store} interface which is described in Sect.~\ref{sect:store}.
The algorithms and collectors are described in Sect.~\ref{sect:algoimp}.

The AGIS system is deployed on a local multi-processor machine dedicated to Gaia. All the
classes are available on each node but objects will be run on specific nodes
according to the configuration
specified in the {\em agis.properties} file. Objects on different hosts communicate through
Remote Method Invocation (RMI), although we actually use  JBoss remote-method calls for efficiency. 
This would be an ideal candidate for Enterprise Java Bean (EJB) implementation but we found EJBs very
inefficient.
In general a class with the name {\em SomeServer}
will only have one instance on the cluster,  while the {\em DataTrain} may have numerous instances, 
e.g., one on each node in the cluster. Internally the {\em DataTrain} makes use of multiple 
processors and cores available in a node.

\section{Data access} \label{sect:dataaccess}

The key to an efficient implementation of AGIS is in the data access.
Even with today's machines, accessing a large volume (tens of terabytes) in
both spatial and temporal order is demanding. 

\subsection{Data access patterns} \label{sect:accpat}

Looking at the four main blocks of AGIS we see that each has a
seemingly unique data access pattern, viz.:

\begin{tabular}{l p{0.65 \textwidth}}
   Source    & All observations of a given source -- spatial\\
  Attitude & All observations within a given time period -- temporal\\
  Calibration & {\raggedright All observations within
  a given time period falling on a given CCD -- temporal/spatial} \\
  Global & All observations -- any order\\
\end{tabular}

\begin{figure}
\begin{center}
\centerline{
\includegraphics[width=0.50\textwidth]{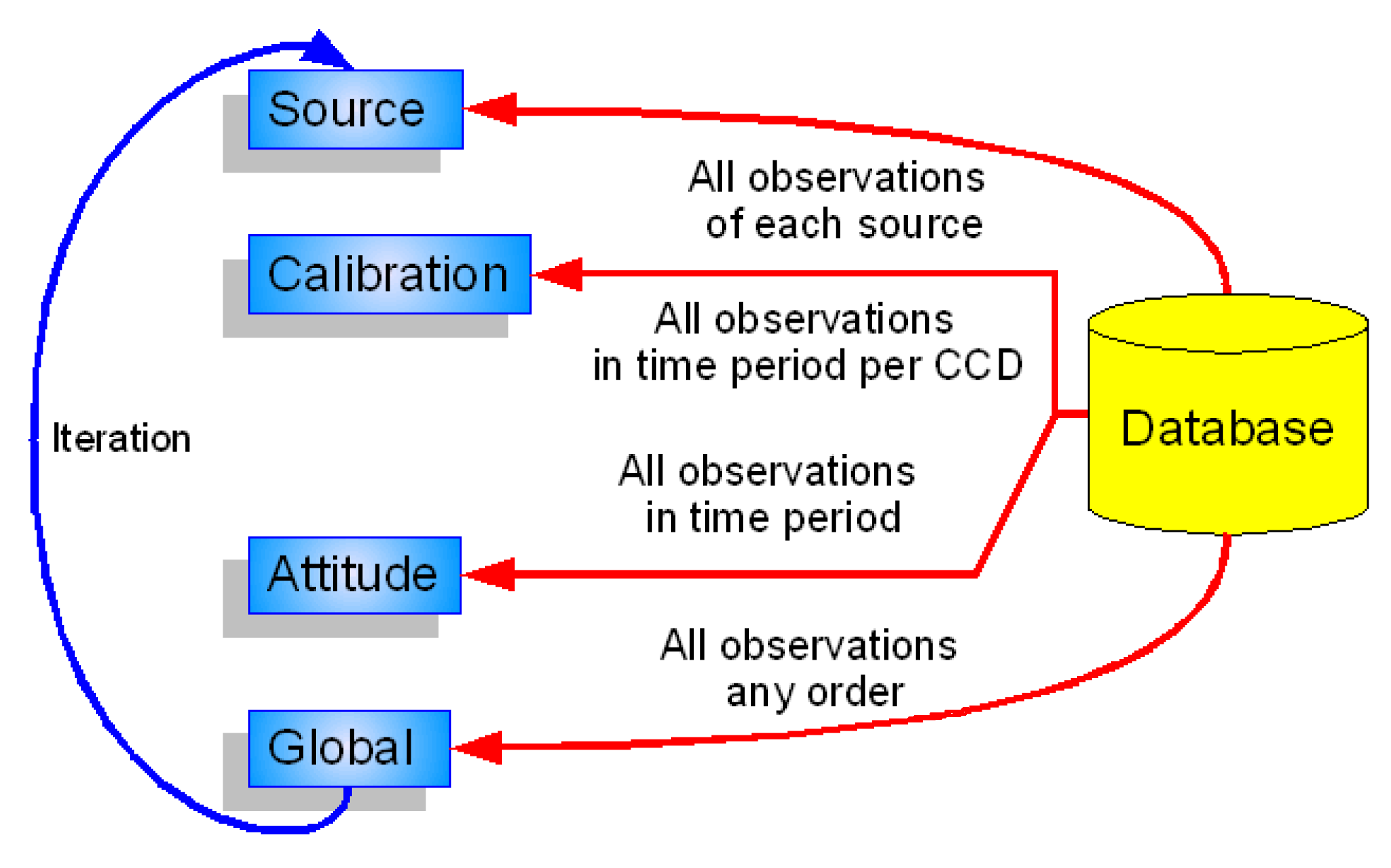} \hspace{0.03\textwidth}
\includegraphics[width=0.47\textwidth]{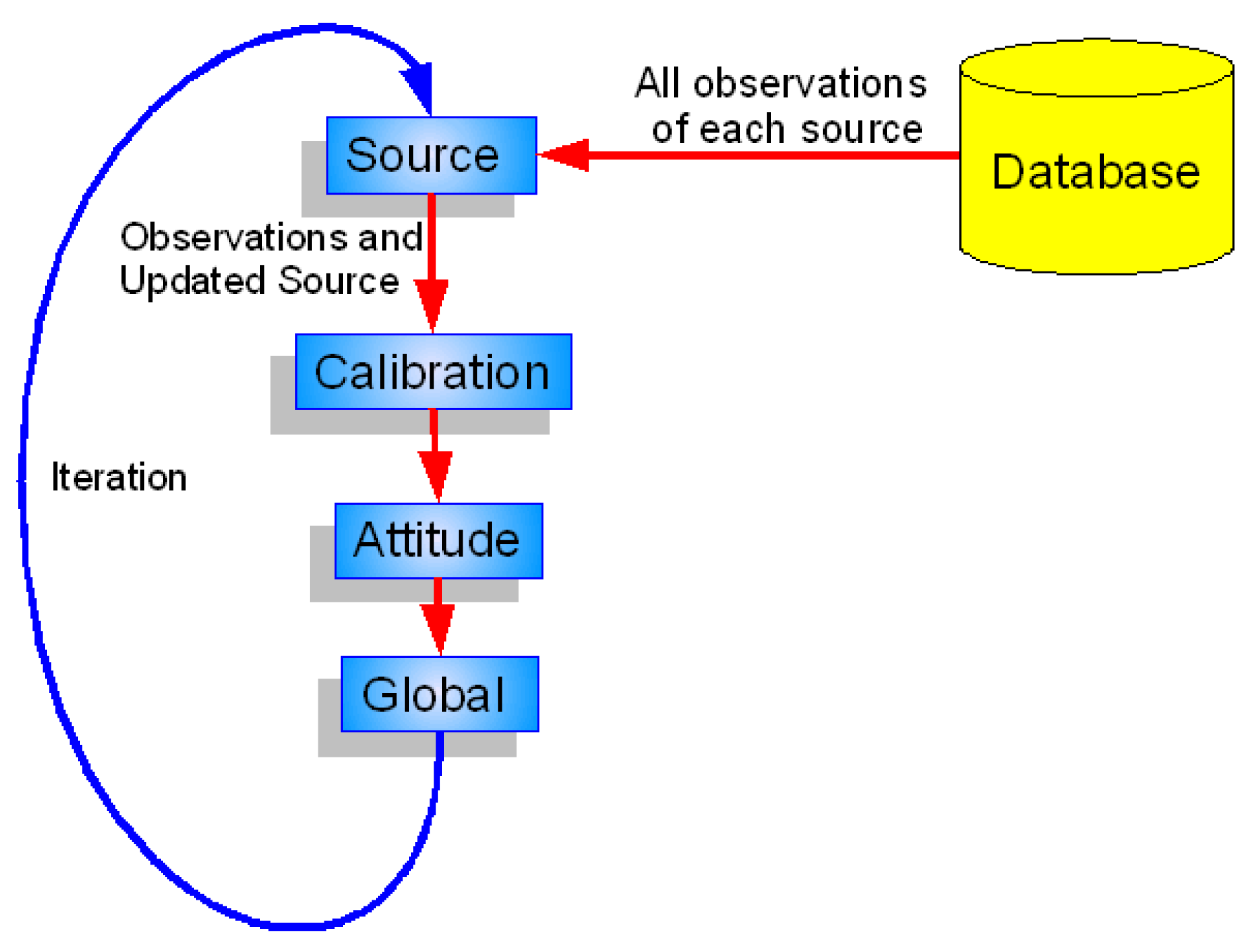}}
\caption{\emph{Left:} Each block of the AGIS solution has a slightly different 
data access requirement. This could cause four passes through the data for each 
AGIS iteration. However, it is immediately
clear that some of these could be combined, e.g., the calibration and attitude
updates could run together, and similarly the source and global updates.
\emph{Right:} With a little in-memory accumulation in the calibration, attitude and global 
update blocks, a complete iteration can be made in one pass though the data. 
Hence the optimal ordering is spatial. Furthermore the updated source parameters
may already be used in the other blocks.}
\label{fig:accessBlocks}
\end{center}
\end{figure}

\noindent
(The `observations' here refer to the {\it AstroElementary} objects described
in Sect.~\ref{sect:dm}.)
The naive approach would be to go through the data once for each block,
updating the parameters in turn and then repeating this for each iteration. This is
indeed the basic mathematical formulation of the block-iterative solution
method and the corresponding data access scheme is depicted in
Fig.~\ref{fig:accessBlocks} (left). 

Running through the approximately ten terabytes of data four times per iteration
is rather daunting, considering that many tens of iterations will be needed. 
Immediately, though, we see that the calibration and attitude updates are similar enough 
that the can perhaps be combined. The global update is order-independent and as such could be
combined with the data access of any of the other blocks, for example source. Indeed
this was already remarked in \cite{1999BaltA...8...57O}, where the prototype made just two
passes through the data for each iteration rather than four.
The question then is: could this be reduced to one pass through the data per
iteration? 

\subsection{A question of order} \label{sect:dataorder}

Let us assume that all four blocks could be executed in one pass; what then would be the
impact of the ordering of the data? There are two primary orderings we may choose: spatial or temporal.


\paragraph{Temporal ordering}
If we assume an ordering based on the time of observation, then for the attitude we may read the data once,  
break it in time chunks suitable for the attitude update, process each chunk in turn
and finish with it. With a
small buffer we may also accumulate the observations required for the calibration and
similarly finish with calibrations in a timely manner during the same pass through
the data. For the global update the order is immaterial, so it can be done in parallel
with the attitude and calibration updates. 

The problem here comes with the source update. Since any given source is observed many times
over the entire mission, if we process in time order we must accumulate the data
for each source until we have all observations of it. This will not
happen until we have seen all of the data -- only then can we be certain that no
more observations of a given source will show up. This would effectively mean that all
observation data would end up in memory. For a hundred million sources  
(with almost $10^{11}$ observations) and 
some clever organizing this would be of the order of 5~TB of data, which
 is infeasible to have in shared memory
on our budget. The final solution may require five times as many  observations.
The alternative is
another pass through the data in spatial order. Since we must wait until the
end of the first pass for the updated calibration, attitude and global parameters,
these updated values could already be used for the source update.


\paragraph{Spatial ordering}
If we assume a spatial ordering, i.e., that all observations of a source are clustered together, 
then the story is quite different. Now we may process each source to find
its new astrometric solution, which can immediately be written out to disk. Since
we are finished with that source, the updated parameters may be used to find its
contributions to the global parameters. The situation for the attitude and calibration 
updates is however that all contributions from all observations must be accumulated until
the end of the pass through the data -- only then may the calibration and
attitude updates be calculated. It is important to note that it is not the
observations which must be held but their contribution to the matrices of
attitude and calibration, which is much smaller than the accumulation of the source
matrices in the temporal ordering. The entire attitude accumulation for the
five year mission data can be done in 8~GB of memory.  The size of the calibration matrix 
depends on the number of calibration artifacts -- currently it requires about 4~GB but 
is estimated to need as much as 32~GB when additional calibration parameters are added in the 
coming years.
Hence with spatial ordering  one pass may be made
though the data for each AGIS iteration, as depicted in Fig. \ref{fig:accessBlocks} (right), 
and a minimum amount of data needs to be held in memory.

This clearly represents a better approach to the ordering from a technical point
of view. Additionally, it is more natural to keep
astronomical data of the same part of the sky together and easily accessible.
Hence the AGIS database has observations of the same source sequentially grouped
together on disk. 

\subsection{Getting data to the algorithms: the {\it DataTrain} and {\it Taker}}
\label{sect:datatrain}

Throughout the Gaia processing there are choices to be made concerning data access
patterns such as those outlined in Sect.~\ref{sect:dataorder}. The ideal approach, for
efficiency, is a data driven approach whereby data is accessed in the sequential
order in which it is stored. Hence rather than algorithms requesting
data they should be presented with data by a {\em mediator}. The
mediator
pattern \citep{book:patterns} is a very powerful tool for decoupling software
modules. The implementation of the mediator for the astrometric solution is
called the {\it ElementaryDataTrain}. 

\begin{figure*}
\begin{center}
\includegraphics[scale=0.35,trim=1cm 0cm 0 0]{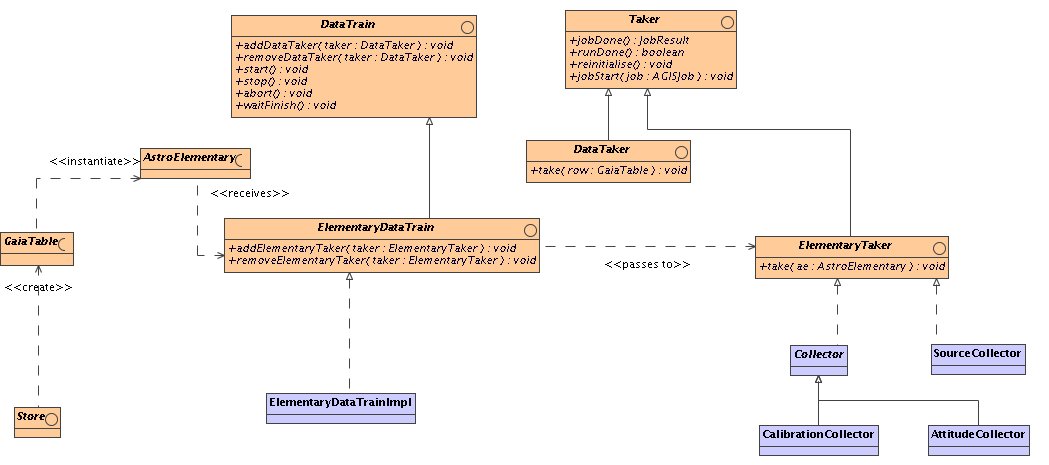}
\caption{The {\it DataTrain} acts as a mediator between algorithms and data
access (the {\it Store}) thus leading to a less coupled system. The {\it ElementaryDataTrain}
accesses {\it AstroElementary}s in the fastest possible manner for the AGIS  
algorithms. The participating algorithms must implement the {\it Taker} interface.}
\label{fig:mediatoruml}
\end{center}
\end{figure*}

The generic notion of a {\it DataTrain} (Fig.~\ref{fig:mediatoruml}) is to
access data in the fastest possible manner, usually meaning
sequentially, and call a given set of algorithms passing them the data. The
concept and code are quite simple. To enable the calling of the algorithms in a
generic manner they must implement the 
{\it Taker} interface, which has a method to `take' some data. By implementing
this interface, the algorithm will have its input when it is called by the {\it DataTrain}.

\begin{figure}
\begin{center}
\includegraphics[scale=0.4,trim=1cm 5cm 1cm 0]{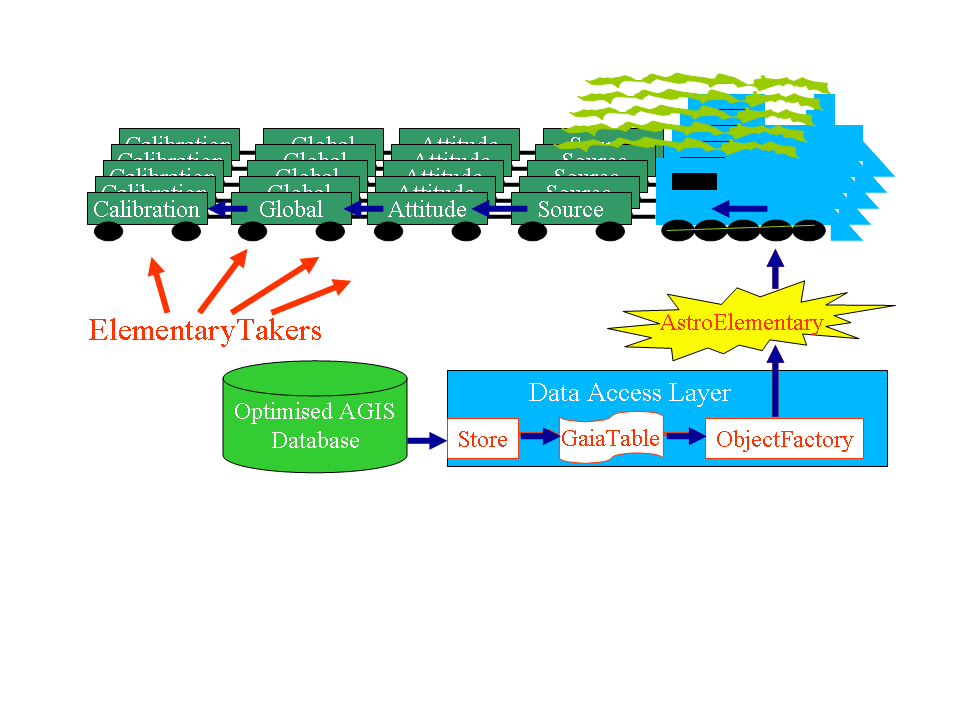}
\caption{Here the blue arrow shows the flow of data from the database through the
{\it Store} and {\it ObjectFactory} to the algorithms attached to the {\it ElementaryDataTrain}.
We may think of the {\it ElementaryDataTrain} as driving through the database, passing
observations to the algorithms.  We may have as many trains in parallel as we
wish.}
\label{fig:datatrain}
\end{center}
\end{figure}

More specifically, for
AGIS the {\it ElementaryDataTrain} accesses {\it AstroElementary} objects,
which are effectively the observations of a given source.
The train decides which data to access by taking a {\it Job}
(see Sect.~\ref{sect:whiteboard}). It uses the {\it Store} to access a set of 
{\it AstroElementary} objects, each of which is then passed to each registered
{\it ElementaryTaker}, i.e., the source, attitude, calibration and global update 
algorithms. Each algorithm (see Sect.~\ref{sect:algoimp}) must implement
the {\it ElementaryTaker}
interface to allow the {\it DataTrain} to interact with it. The
{\it ElementaryDataTrain} has a method for registering the algorithms ({\em addElementaryTaker} 
in Fig.~\ref{fig:mediatoruml}). The algorithms must then accumulate
observations until they can process a particular source or time interval. This
forces the algorithms to accept data in the order it
is stored allowing the infrastructure to be built without fixing the data
storage order. Choosing spatial ordering (Sect.~\ref{sect:dataorder}) means that
all of the elementaries for a given source are sequential. Any given train accesses complete
sets of elementaries with respect to sources. The cartoon in Fig.~\ref{fig:datatrain}
depicts this in a another manner showing how the {\it AstroElementary} is
constructed by the {\it ObjectFactory} from a {\it GaiaTable} resulting from
a query to the database through the {\it Store} interface. The
{\it AstroElementary} is then passed to the algorithms attached to the {\it DataTrain}.

\subsection{Abstraction of data storage: the {\it Store}} \label{sect:store}

To give a degree of independence from the physical storage mechanism, it is
normal to use some abstraction. Java interfaces provide an excellent approach to
provide such insulation. Creating an interface is a small coding overhead,
while in usage one gets a real implementation, i.e., without overhead. It is
very important to realize that a Java interface is a contract binding the using
class and the providing class but does no translation of any kind.  
This should not be confused with rooted persistence systems requiring all classes to inherit 
from some root class. Here we simply have to implement a few methods implied by the interface. They 
are  more for our convenience than a design principle -- we also like to keep clear in our code which 
objects we will be storing and which we will not. It is also useful in the \class{ObjectFactory} to have a base 
interface to cast to, other than \class{Object}.
We are not far from Java Persistence 
Architecture (JPA) in both principle and implementation -- this however was not mature when we
started in 2005. More recently we have considered simply switching to something like Hibernate but 
found the offerings far slower than our own system. We could remove the restriction of having
\class{GaiaRoot} but it has not been a pressing issue to date. Having our own \class{Store}
also  made it easy to build a \class{CacheStore} which took advantage of the Cache high speed interface,
they do not support JPA.

\begin{figure*}
\begin{center}
\includegraphics[scale=0.35,trim=1cm 0cm 0 0]{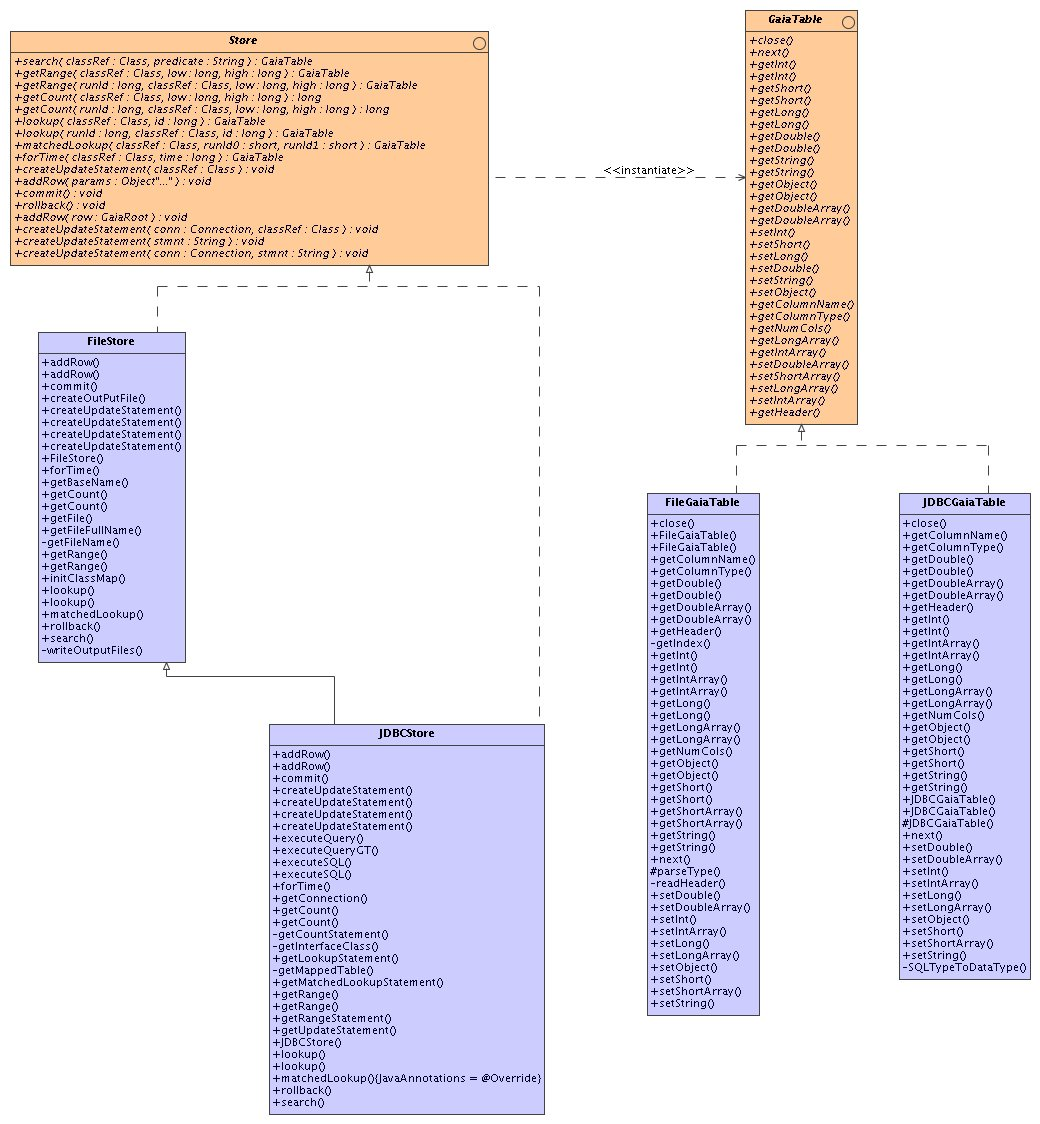}
\caption{The {\it Store} provides an interface for data access, whereby  
many {\it Store} implementations may exist. In the Figure we see a {\it FileStore}
and a {\it JDBCStore}, both of which implement {\it Store}. With these
implementations of AGIS 
code we may switch between FITS files and a JDBC Database for storage in a
seamless manner.}
\label{fig:storeuml}
\end{center}
\end{figure*}

No algorithm code in the system interacts directly with the database
management system; rather a query interface to the data is provided through
the {\it Store} interface (see Fig.~\ref{fig:storeuml}). The implementation
of the {\it Store} is hidden behind the interface; thus the data store may be
implemented in files or any database management system.

An implementation of the {\it Store} is requested from the {\it AGISFactory}, the actual implementation of
the store is configured using the {\it gaia.tools.dal.Store} property in the
{\em agis.properties} file and
thus can be changed at run-time (rather than at compile-time). The {\it Store} interface includes an explicit
range query which returns all objects within a certain id range, which is
required to support the {\it DataTrain}.

As depicted in Fig.~\ref{fig:storeuml} there are multiple implementations of the
{\it Store}. The {\it FileStore} does not support the same level of querying
as the {\it JDBCStore} but is sufficient for running the testbed on a laptop.  Most
recently we have also implemented a {\it CacheStore} over the InterSystems Cach{\'e} database.

{\it GaiaTable} in Fig.~\ref{fig:storeuml} represents an interface to tabular data.
The assumption of dealing only with tabular data is a major simplification for AGIS. This is a
fair assumption dealing with astrometry data. Both files (be they
FITS or whatever) and relational database tables may be represented as a {\it GaiaTable}. 
The interface defines methods for retrieving the next row and for getting columns by name
or index. The whole row may be passed to the algorithm or
{\it ObjectFactory} and it may extract the required columns. The
{\it DataTrain} loads the entire row.

\begin{figure*}
\begin{center}
\includegraphics[scale=0.3,trim=0 0cm 0 0]{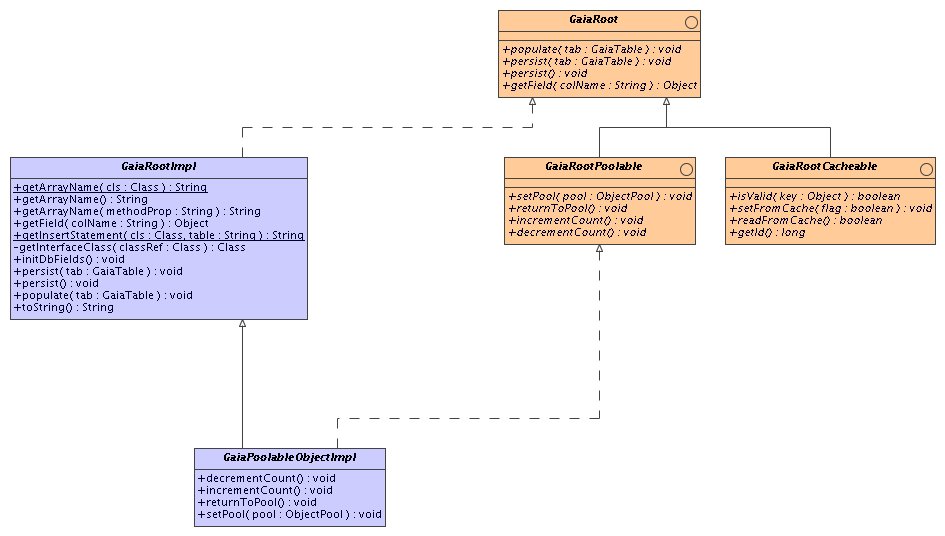}
\caption{All data objects implement {\it GaiaRoot}, which makes certain methods
available to the {\it Store}. All data objects are interfaces, not real classes -- this
allows them to be easily replaced by different implementations.}
\label{fig:gaiaroot}
\end{center}
\end{figure*}

The {\it GaiaRoot} UML (Unified Modeling Language) diagram is given in Fig.~\ref{fig:gaiaroot}. 
Color interfaces are 
shown in brown colour (and are also marked with a $\circ$), while
implementations are in blue. Any objects in the Gaia data model which use the
{\it Store} (see also Fig.~\ref{fig:storeuml}) and {\it ObjectFactory} must
implement this interface. A basic implementation is provided which most classes
may inherit from, but in some cases, due to single inheritance in Java, this may
not be possible. In fact practically all of the required functionality is in the 
{\it Store} or {\it ObjectFactory}. 

Interfaces were chosen for the data classes originally, since the first implementations
in 1998 used Objectivity/DB  (from Objectivity, Inc.) which was a rooted system, thus requiring the objects to 
actually inherit from the Objectivity/DB base class. Even then the {\it Store} was working both with Oracle
Real Application Clusters
and Objectivity/DB, which meant having two implementations of the data objects.  These 
days we usually only have one implementation; however, there are instances where the interfaces 
are still useful. For example higher-level classes such as {\em AstrometricSource} can have multiple
subclasses. These may not follow the same inheritance hierarchy but can still be {\em AstrometricSource}s 
since it is an interface; if it were only a class there could be inheritance problems.

\subsection {Access to objects: the {\it ObjectFactory}\label{sect:objectfactory} }

The {\it Store} deals essentially with tables but some code will require objects. The
{\it ObjectFactory} sits on top of the {\it Store} and returns objects
implementing  the data model
interfaces.  The object-from-table method of the interface is also exposed, allowing 
code to do this conversion exactly when required.
We need to take care that not too many pieces of code perform such a
transformation -- preferably it would be done once by the {\it DataTrain}.
Splitting this out allows for very direct measurement of the performance.

This is implemented as a {\em Generic} class. The {\it Factory} is instantiated for a
specific data model interface and then provides a method returning that class of object only. Java
Generics are very nice for this and, although similar to C++
templates, should not be confused as being the same. Generics provide type
checking and safety but they do not generate extra code with new types.

The {\it Factory} relies on the {\em populate} method of the {\it GaiaRoot} to
populate the fields of the object from a {\it GaiaTable}. A generic implementation of this using a
mapping from the configuration file is provided in the {\it GaiaRootImpl} class. This provides a convenient 
mechanism to read the data from the {\it Store} 
into a Java object that can be used
elsewhere in the system.

The {\it ObjectFactory} also has caching capabilities. Whenever an object is read from the {\it Store} 
it may be cached in memory in order to avoid new reads when it is requested again. 
Any object which is created by the {\it ObjectFactory} can be made cacheable just by
implementing the {\it gaia.tools.dm.GaiaRootCacheable} interface. The caching can also be 
disabled by adding an entry to the
property file. The interface contains a method to determine the `validity' of the object. 

The {\it Factory} also has the possibility to implement object pooling. The notion here
is to reuse objects by filling them with new data rather than reconstructing new
ones. This technique was very popular in early Java implementations to reduce
garbage collection time. Tests with the new JDK (1.5 and 1.6) show that this is no
longer beneficial. Still, by having all data object creation done through one
class the possibility to change the way it works later remains available.

\section{The Data Model}\label{sect:dm}

The algorithms work in terms of
Java objects such as {\it Source} and {\it AstroElementary}.  These objects
form what is generally termed a Data Model for the system.  

The data used for AGIS will comprise between 10\% to 50\% of the sources 
(and their corresponding observations), corresponding to the so-called primary
sources briefly discussed in Sect.~\ref{sec:secondary}. The selection of the
primary sources is described in \citet{LL08} and is implemented as several database 
queries. The selected data will be put in the special AGIS database (see for example 
Fig.~\ref{fig:datatrain}). 

The data model is made in terms of
interfaces to allow easy substitution of multiple implementations. 
The {\it ObjectFactory} (Sect.~\ref{sect:objectfactory}) and
{\it Store} (Sect.~ref{sect:store}) are used to construct real implementations
of these interfaces but all code refers only to the interface. Hence all client
code may be compiled without any implementation if necessary. This is a technique
used throughout AGIS and indeed also for {\it GaiaTools}, the common software toolbox 
for all Gaia processing tasks. 
The most important interfaces are:
\begin{itemize}
\item \class{AstroElementary}: 
An object of this kind represents the transits of a celestial source over  the
first dedicated 10 CCD strips of the focal plane, namely, SM1 or SM2 and AF1--9
\citep[see][for an outline of the CCDs in the focal plane]{LL08}.
Each {\it AstroElementary} in AGIS is uniquely associated with a {\it Source}.

\item \class{Source}: 
An object of this kind represents celestial sources that follow the  
standard astrometric model (thus modelled by the six astrometric parameters
described in Sect.~\ref{sec:intromath}) and are eligible for AGIS source processing.

\item \class{Attitude}:
An object of this kind represents an interval of continuous attitude data.
Attitude is parametrized using B-spline coefficients of a given order representing 
the four components of the attitude quaternion (Sect.~\ref{sect:attupd}).

\item \class{CalibrationEffect}: 
The geometrical calibration of the instrument is made up of multiple 
\class{CalibrationEffect}s   
(Sect.~\ref{sect:calupd}) all of which may be configured in an XML file.
\end{itemize}

\begin{table}
\caption{Evolution of AGIS performance during the development of the processing
framework. Data volumes are indicated by the number of observations ({\it AstroElementaries}),
depending on the number of sources and the length of the observation period. The time is the
processing time per AGIS iteration for the given number of processors. The last column
shows the throughput, in observations per processor per hour, as an indication of the real 
performance.}
\label{tab:agistimes}
\begin {center}
\begin{tabular}{ccccc}
\hline\noalign{\smallskip}
Date & Observations & Processors & Time (hr)  &  Normalized Rate\\
\noalign{\smallskip}\hline\noalign{\smallskip}
2005 & $1.6 \times 10^7$  & 12 & 3 & $0.9 \times 10^6$ \\
2006 & $8.0 \times 10^7$  & 36 & 5 & $0.5 \times 10^6$ \\
2007 & $8.0 \times 10^7$  & 24 & 3 & $1.3 \times 10^6$ \\
2008 & $8.0 \times 10^7$  & 31 & 1 & $3.2 \times 10^6$ \\
2009 & $2.6 \times 10^8$  & 50 & 1.8 &$2.8 \times 10^6$ \\
2010 & $4.0 \times 10^9$  & 68 & 9.5 &  $6.2 \times 10^6$ \\
\noalign{\smallskip}\hline
\end{tabular}
\end{center}
\end{table}

\section {Distributed processing } \label{sect:distproc}

Regardless of the ordering chosen (Sect.~\ref{sect:dataorder}) the access of the data
does not need to be done serially. Indeed we require the data to be sequential
on disk but multiple parts of that sequence may be read simultaneously. In the
case of sources we may process simultaneously each source, in terms of
distributed computing this is `embarrassingly parallel' \citet{book:pprog}.%
\footnote{\url{http://en.wikipedia.org/wiki/Embarrassingly_parallel}}
We may theoretically gain up to a factor $N$ in speed by using $N$ processors, if $N$ is
the number of sources to be processed. We say theoretically with reason, as the
data must still be read from disk and we are unlikely to actually put in place
$10^8$ processors. Still, tests have shown that the processing time indeed 
decreases in proportion to the number of processors used for AGIS. 
Some numbers are given in Table~\ref{tab:agistimes}.


\subsection {Distributed processing frameworks} \label{sect:distframe}
Many different approaches exist for distributed processing, and they are 
usually embodied in some library.
However since we have an `embarrassingly parallel' problem we have little need for such a 
complex and heavy library. In fact  all that we require is already available
within the standard Java library, namely:
\begin{itemize}
\item Communication between processing nodes: the Remote Method Invocation (RMI) framework in Java provides this.
\item Access to a database or databases: the Java Data Base Connectivity  (JDBC) framework provides this.
\item Some form of graphics library for GUIs: Java Swing library provides this. 
\end{itemize}
Additionally, in this age of the web, Java provides easy support for dynamic web site 
generation using Java Server Pages (JSP).

Hence an early feeling was to use the tools of Java directly, rather than try to
fit the problem into one of the many distributed programming 
libraries, each with their own assumptions and problems. The modern programming
languages of the day, such as Java, are very sophisticated in the feature set and
tools they provide. For example the Java/Jini Parallel Framework \citep[JJPF;][]{danelutto2005} 
provides
some reliability on top of these tools while also taking a much more process-oriented 
view -- each worker has a {\em getData} call to pass back results. JJPF
is also more coordinator oriented with a single server eliciting support from
available nodes to perform a computation. In the grid world the obvious
contender would be the Globus Toolkit \citep[GTK;][]{foster2006}. Previous forays into
GTK showed the system to be buggy and difficult to use. GTK has
improved dramatically over the years, yet it still remains service oriented 
(we believe our problem to be data oriented) and has a large security overhead which
we do not see as necessary. Indeed though \citet{2003cs.......11009D} is positive
about GTK they introduce a resource broker which seems similar to our whiteboard
(Sect.~\ref{sect:whiteboard}). Unfortunately \citet{} say little
about the data intensive applications mentioned in the title of their paper.
 
The notion of just using the Java framework without some other layer was
reinforced by previous experiences with
the Sloan Digital Sky Survey (SDSS).  On the SDSS a form of distributed query
system known as CasJobs \citep{conf:casjobs} was built using Web Services, 
the SQLServer database and the C\# language.%
\footnote{The \# here is the musical sharp; hence this is pronounced `See sharp'.}
This was done
quite rapidly without using any special libraries beyond the facilities 
available in the programming language.  Within the same group at
Johns Hopkins a typical Grid application for finding galaxy clusters in a
large catalogue was taken and quickly rewritten in C\#. As reported by 
\citet{2005cs........2018N}, this ran about ten times faster using a database 
system than the traditional file-based Grid system.

A final justification, perhaps the ultimate and obvious one, for not taking on a
library is that of simplicity. It was believed the distributed computing
libraries would not make the system simpler hence none were adopted.

\subsection{Job distribution: the {\it Whiteboard}} \label{sect:whiteboard}
There are at least two main approaches to controlling a grid of distributed
processes. The first is to have `agents' register with some central controller
which then regulates the entire process; the second is to have a less centralized
approach with more autonomous processes. 

The central controller approach is very appealing and generally the way many
agent-based systems work. Generally these involve monitoring resources and
farming out jobs to particular processors which are not fully loaded. The central
registering of agents means the controller knows how many agents of which types
exist on the system, and furthermore may reject agents from particular machines
or of particular types. Such systems deal well with uneven workloads and ad hoc
jobs by many users. Often security layers and user tracking are included.

By contrast, an AGIS iteration could easily occupy an entire cluster for some
days. There are no ad hoc programs, only the entire AGIS chain running on all
data. There are no users, hence no particular need for a security overhead in
terms of certificates, etc. 

\begin{figure}
\begin{center}
\includegraphics[scale=0.45,trim=0cm 0cm 0 0 ]{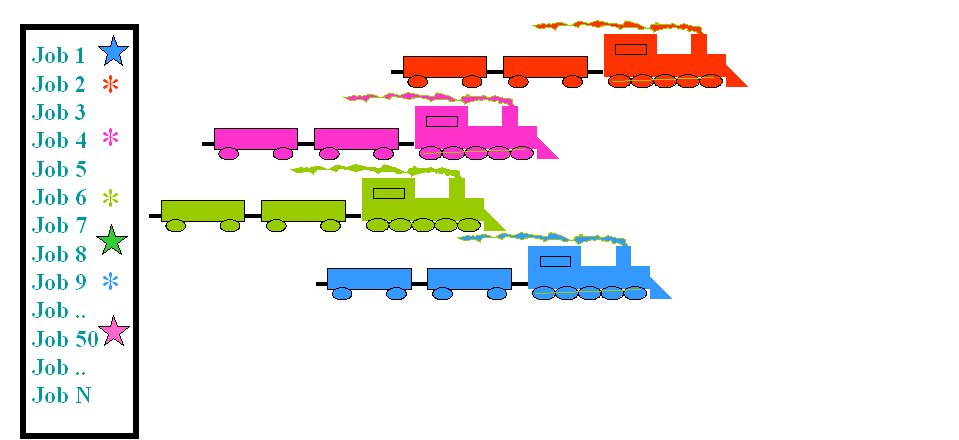}
\caption{A set of jobs corresponding to sequential batches of data which
cover the entire data range may be posted on a
whiteboard (left). The {\it DataTrain} marks a job as in progress when it starts it
and as completed when it is finished. There may be many {\it DataTrain}s (right).
When all jobs are done all of the data have
been seen once. The whiteboard itself has no special knowledge of the
jobs or the overall task -- it is a simple mechanism to coordinate potentially
hundreds of processes.}
\label{fig:whiteboard}
\end{center}
\end{figure}

In our data driven approach (Sect.~\ref{sect:datatrain}) we may consider the data as
the distribution mechanism. Everything hinges on the processing of some
block of data, be it a time sequence or a set of spatially ordered
observations (Sect.~\ref{sect:dataorder}). All we really need to know is if a
particular part of the data set has been visited during a particular iteration.
If the data segments are chosen properly we may have as many {\it DataTrains}
running as we wish.
This is very simple and easily achieved through a whiteboard
mechanism as depicted in Fig.~\ref{fig:whiteboard}.

\begin{figure}
\begin{center}
\includegraphics[scale=0.45]{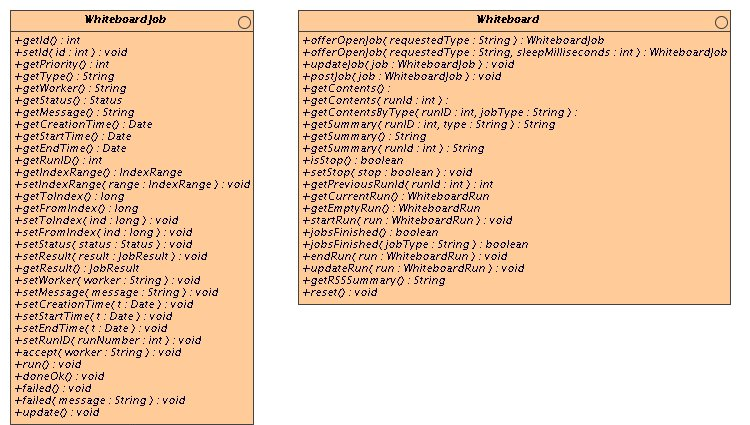}
\caption{The UML interfaces for the {\it Whiteboard} and the {\it WhiteboardJob}. Note the
{\it postJob} method used to populate the whiteboard and the {\it offerOpenJob} methods which the
{\it DataTrain}s use to get jobs. The job itself has methods for status and messages, etc.}
\label{fig:umlwhiteboard}
\end{center}
\end{figure}

The whiteboard is quite a simple concept for organizing many processes of
varying types. Conceptually we may `post' jobs on a whiteboard, and then workers, 
in our case {\it DataTrains}, may pick them up. The whiteboard may hold status
information, e.g., about when a job started, when it ended, if all was OK, etc. In effect
then the whiteboard becomes the central controller, although it exercises no control
as such. Perhaps the original of the species in this respect is the OPUS
pipeline from the Space Telescope Science Institute \citep{1995ASPC...77..429R}.
Indeed, it is the OPUS blackboard%
\footnote{Whiteboard was elected as a more modern alternative to Blackboard.} 
design pattern which is employed here. It is
noted that since its early beginning, OPUS is itself moving toward Java
\citep{2003ASPC..295..261M} but maintaining its heterogeneity through CORBA (Common
Object Request Broker Architecture). For the purposes of AGIS, which is a pure
Java implementation, a simple {\it Whiteboard} was coded directly in Java using a
database table to hold the jobs. The latter also provides the ability to ensure
that no two trains ever get the same job. The JDBC framework in Java makes the
whiteboard seamlessly accessible from any node on the network -- hence no need
for the overhead of CORBA or some other message passing system here. The UML
interface for the {\it Whiteboard} is shown in Fig.~\ref{fig:umlwhiteboard}. 

Regardless of the jobs being done, the whiteboard can give some information on
the general state of the system. A Series of JSP pages present the whiteboard
state on a website. On this site with little effort we may show jobs
completed/remaining and (assuming uniform jobs) an estimate for the end time. 
We may also list statistics per processor simply by querying the job table
in the database.

\subsection{Overall control: the {\it RunManager} and {\it ConvergenceMonitor}} 

\begin{figure}
\begin{center}
\includegraphics[scale=0.45]{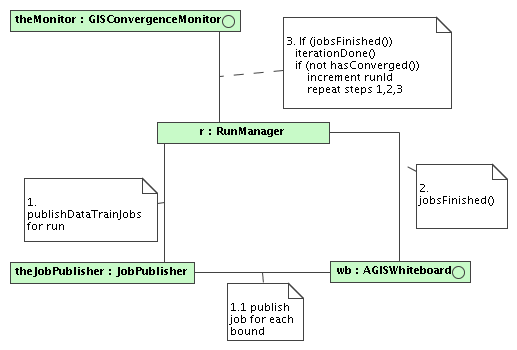}
\caption{Communication diagram for the  {\it RunManager}. This summarises the
{\it RunManager}s role in publishing jobs and checking for convergence.}
\label{fig:runmanager}
\end{center}
\end{figure} 

The {\it Whiteboard} alone is not enough to run an AGIS
solution. Some other entity must post the jobs on the board for the
{\it DataTrains} to work on. The {\it RunManager} has the task of 
coordinating iterations and the publishing of jobs as depicted in
Fig.~\ref{fig:runmanager}. The {\it RunManager} uses the {\it JobPublisher} to
publish appropriate jobs, e.g., one for each block of sources to be processed.
The {\it JobPublisher} scans a table of bounds (a list of identifiers  of
elementaries which are the last
in a series belonging to a single source) and creates a number of jobs based on
blocks of elementaries. In general the system is configured such that these
jobs complete in a few minutes, as this gives a better indication of progress and
the need to redo a job, in case of a problem, is detected in a timely manner. 
Hence there are typically thousands of jobs in a single run. Once posted, the
trains pick them up and start working. The order in which the jobs are done does
not matter.  Jobs are also published for the calibration, attitude
and global updates if these algorithms are attached to the train. These jobs execute 
for the entire iteration.

The {\it RunManager} then periodically checks to see if the {\it DataTrain} jobs
have finished. If they are done the main part of the iteration is done, and the
{\it GisConvergenceMonitor} is told the iteration is at an end. The
{\it RunManager} then asks the {\it GisConvergenceMonitor} if the solution
has converged and awaits the answer. At this point
the attitude, global and calibration servers still must perform their final
calculations -- when these are complete the {\it GisConvergenceMonitor} reports
the state of convergence. The convergence criterion is currently based on the 
typical size of the source updates in the current iteration. 

If convergence has not been reached the {\it RunManager} starts another run
through the data by publishing a new set of jobs. If it has converged the {\it RunManager}
declares the run ended and converged.

\section{Algorithms}\label{sect:algoimp}
There are effectively two types of algorithm in the system: those with a
centralized part and those which are completely distributable. Let us first look
at the source update algorithm which is completely distributed and
subsequently at the others. The mathematical formulation of the algorithms (or
blocks) is given in \citet{LL08}. As explained in Sect.~\ref{sect:iteration} 
the blocks are iterated until the solution is considered converged.

\subsection{Source update} \label{sect:sourceupd}

The mathematical details of the source update are provided in \citet{LL08}. Very briefly,
the update for source $i$ is obtained by solving the overdetermined system of equations
\begin{equation}
\vec{A}_i \vec{d}_i \simeq \vec{h}_i \, ,
\label{eq:lsobsmatrix}
\end{equation}
where $\vec{d}_i$ is the $n$-vector of updates to the astrometric parameters
$\vec{s}_i$ of the source (usually with $n=5$, as described in Sect.~\ref{sec:intromath}),
$\vec{h}_i$ the $m$-vector of residuals, where $m \gg n$ is the number of 
observations of the source, and $\vec{A}_i$ the design matrix. The problem is
complemented by an $m$-vector of measurement uncertainties, $\vec{\sigma}_i$.
The residual vector $\vec{h}_i$ contains the observed minus the calculated values 
for the source, such that the $j$th element is 
$h_j=g_j^\text{obs}-g_j^\text{calc}(\vec{s}_i,\vec{n})$, $j=1\dots m$, 
where $g_j^\text{obs}$ is the observed position of the source on the CCD and 
$g_j^\text{calc}$ the calculated position based on the current best estimate of 
the source parameters $\vec{s}_i$ as well as the attitude, calibration and global 
parameters in $\vec{n}$, cf.\ Eq.~(\ref{eq:generalobs}). 
The elements of $\vec{A}_i$ are $A_{jk}=\partial g_j^\text{calc}/\partial s_k$ for
$j=1\dots m$ and $k=1\dots n$. Each {\it AstroElementary}, consisting of up to
10 CCD transits, generates several rows of design equations. 

The least-squares solution of Eq.~(\ref{eq:lsobsmatrix}) is by itself an iterative process,
in order to have a self-adapting system of observation weighting (essentially by
adjusting $\vec{\sigma}_i$) that is robust against outliers. Typically three or four 
such internal iterations are needed to compute the update $\vec{d}_i$, after which
the improved source parameters are obtained as $\vec{s}_i+\vec{d}_i$. The solution
of Eq.~(\ref{eq:lsobsmatrix}) is done in a very standard fashion by forming normal 
equations \citep{book:bjork-1996}, which is computationally very efficient. 

\begin{figure*}
\begin{center}
\includegraphics[scale=0.4]{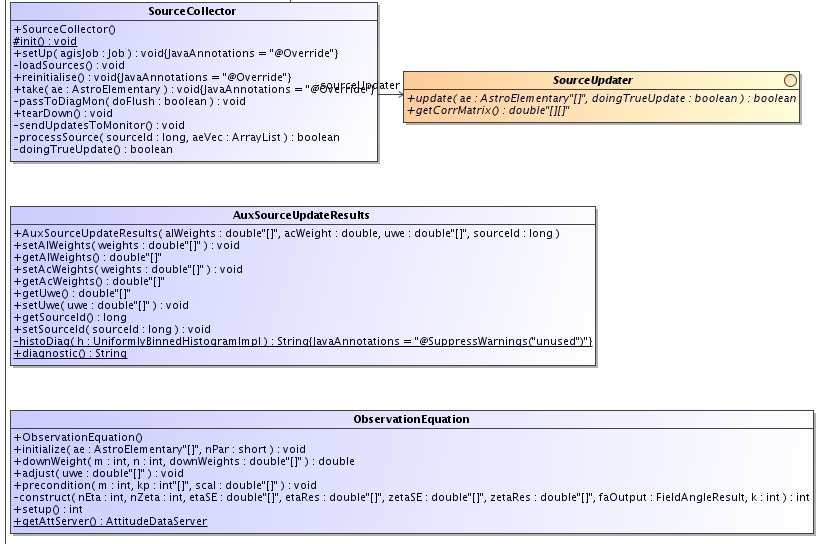}
\end{center}
\caption{\label{fig:sourceUpd} The {\em SourceCollector} is attached to each {\em DataTrain} 
and has a {\em SourceUpdater} associated with it to update all sources on disk when a job is finished.}
\end{figure*} 

The source update step is truly distributed. As the
{\it DataTrain} passes elementaries to the {\it SourceCollector} (the {\it Taker}  
registered with the train for sources)  it accumulates all of the elementaries  for a given source. 
Remember that the data are stored in such a manner that all elementaries for one
source are consecutive; hence, when the {\em sourceId} changes, the collector knows that 
it has all the data for a given source. Once it has a batch of elementaries, the source 
update is called to compute the required update of the astrometric parameters. 

Figure~\ref{fig:sourceUpd} provides a UML overview of some of the classes involved 
in the source update.
When the updated astrometric parameters are available they are passed to the
{\it SourceUpdateManager}, which batches together several sources for efficient
storage. Nothing in AGIS is ever actually updated, rather a new source row is
written to the table with the current {\it runId}. In this way a complete history of
the updates are preserved. Inserting to the database is also more efficient than updating. 

In fact the {\it SourceUpdateManager} does not write the sources finally until the entire 
job is done. When all sources are updated a database transaction is opened to write all
the results -- only when this is done is the job considered finished. 
In this manner a job is either completed or not,
since the transaction may be `rolled back' without consequence if there is some problem.

\begin{figure*}
\begin{center}
\includegraphics[scale=0.4]{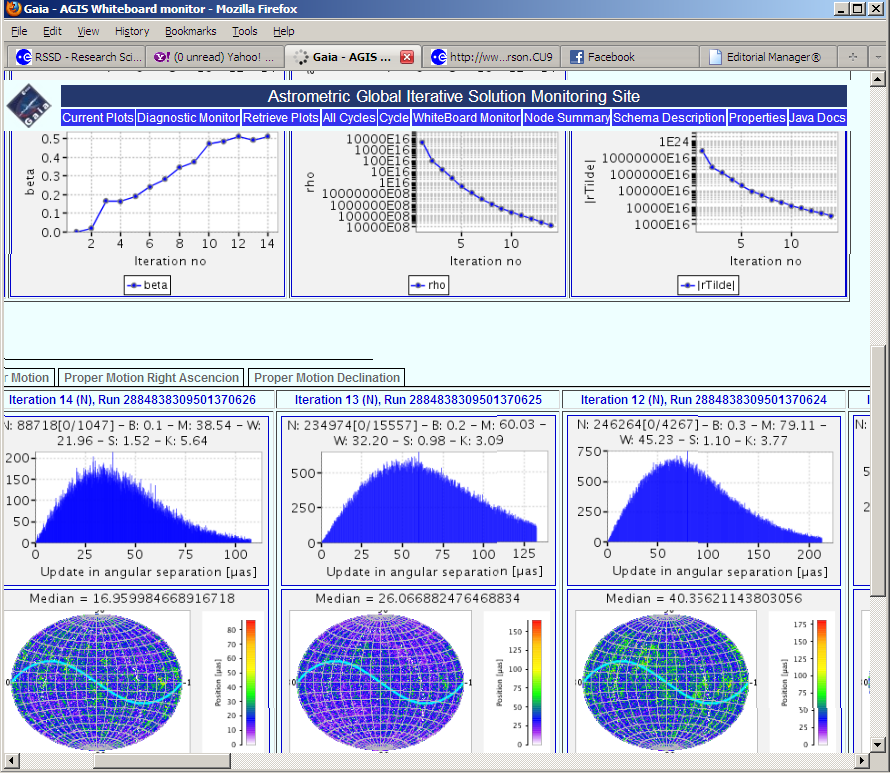}
\end{center}
\caption{\label{fig:agismon} Update plots such as for the source position update
shown here are generated dynamically and displayed on the AGIS monitoring
website while the system is running. The Conjugate Gradients parameters such as 
$\rho$ and $|\vec{\tilde{r}}|$ \citep[see][for details]{LL08}
are also tracked as shown.
Historical plots may  be retrieved from the system.}
\end{figure*}

When the job is finished the {\it SourceCollector} sends all of the updated 
sources to the {\it GisConvergenceMonitor}. This call is made using RMI. Because 
the {\it GisConvergenceMonitor} receives sources throughout the iteration, histograms 
of the updates can be dynamically generated. These are displayed on the 
associated AGIS website in real time. An example of the website is shown in 
Fig.~\ref{fig:agismon}.

\subsection {Attitude update}\label{sect:attupd}

\begin{figure*}
\begin{center}
\includegraphics[scale=0.7,trim=0cm 13cm 0 0]{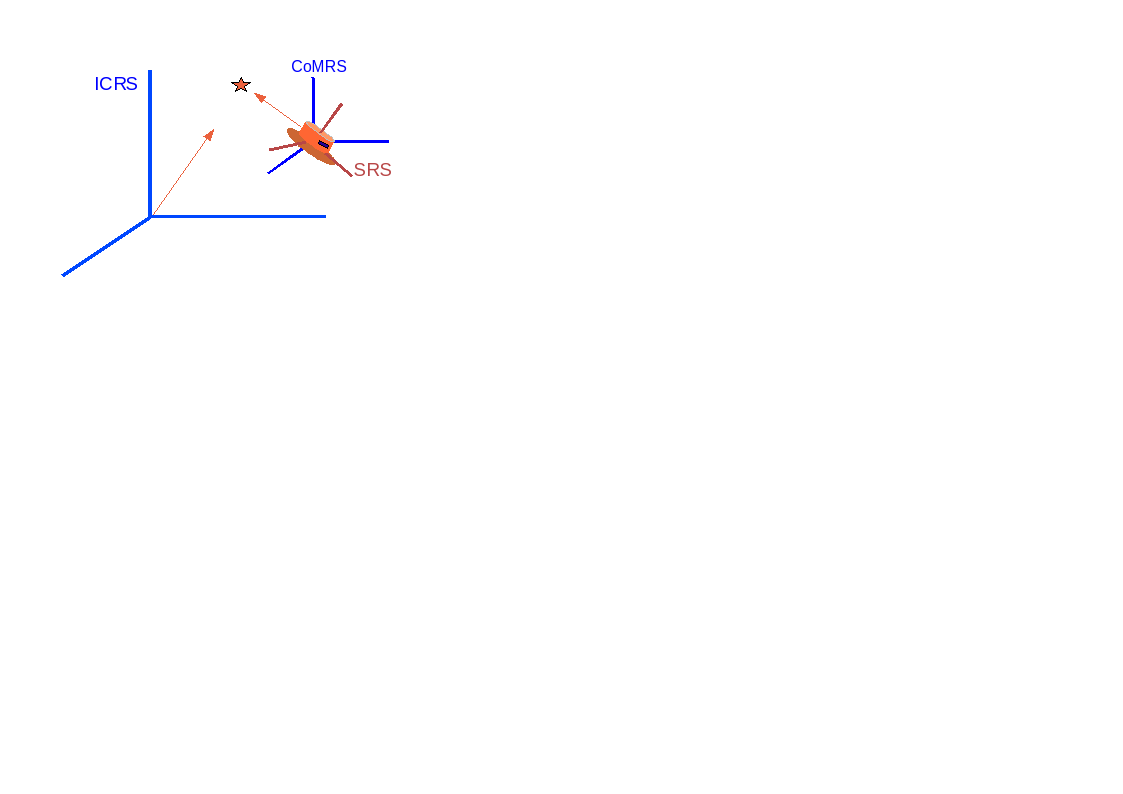}
\caption{ \label{fig:atttoon} The position of a source is initially given in the ICRS, centred on
the solar-system barycentre, and is then transformed to the Gaia-centred CoMRS by taking into
account parallax and relativistic effects. Finally the position may be 
transformed to the SRS (Scanning Reference System), which is fixed to the spinning instrument, 
by means of a rotation given by the attitude quaternion $\vec{q}$.}
\end{center}
\end{figure*}

The attitude specifies the instantaneous orientation of Gaia in the same celestial reference frame 
as used for the astrometric parameters -- for Gaia this is known as the Center-of-Mass Reference System 
(CoMRS). Being the local rest frame of Gaia, the axes of the CoMRS are aligned with the 
International Celestial Reference System \citep[ICRS;][]{icrs1998}, but with Gaia as the origin 
of the coordinate system instead of the solar system barycentre (Fig.~\ref{fig:atttoon}).
While the CoMRS is an inertial frame, the Scanning Reference System (SRS) rotates with the satellite
and the optical axes of the astrometric telescope are fixed in the SRS. To a first approximation,
the CCDs therefore measure the positions of the sources in the SRS. 

The CoMRS and SRS frames are related by a purely spatial rotation, which defines the
instantaneous attitude of Gaia. 
We use quaternions \citep{hamilton1844} to represent the attitude, as  
is common practice for spacecraft \citep[e.g.,][]{kane:1983}. The quaternion ${\bf q}$ is
a 4-vector representing a direction in space (expressed in either the CoMRS or the SRS) and an 
angle of rotation around that direction. The four elements of the quaternion are continuous 
functions of time, here denoted $q^k(t)$, $k=1\dots 4$, which allow a singular-free 
attitude representation for arbitrary rotations. 
These functions are modelled as cubic splines, using short-range B-splines $B_n(t)$ as basis 
functions \citep{book:DB-01}; thus
\begin{equation}
        q^k(t) = \sum_n a_n^k B_n(t) \, , 
        \label{eq:att1}
\end{equation}
where $a_n^k$ are the attitude parameters, of which there are a few million in the system
\citet[see][for details]{LL08}.
The attitude update solves a linearised least-squares problem similar to Eq.~(\ref{eq:lsobsmatrix})
but with the unknowns $\vec{d}$ now being the updates to the attitude parameters $a_n^k$
and the partial derivatives in $\vec{A}$ being taken with respect to these attitude parameters.
The dimension $m$ is however very much greater in this case, since the attitude update
in principle has to consider all the observations throughout the mission. However, thanks
to the short range of the B-splines, the attitude normal matrix is band-diagonal, and the resulting
system can be stored and solved very efficiently. 

In fact the attitude may be divided into segments
each of which can be solved simultaneously but separately. There will  be natural breaks in
Gaia's attitude that can be used to segment the data, but this technique may be used to distribute the
attitude processing further. Hence, depending on the number of attitude  segments, there is
a limit to the distribution of attitude processing. Each segment  may be solved
on an individual processor. In actual fact the final fitting of the attitude
for five years data as a single spline with knots every fifteen  seconds took
only 30 minutes on a Xeon processor with 16~GB of memory.  The
solution itself is not the bottleneck, but rather the gathering of the observations.  With
a single attitude update server all source observations must be passed  to
this server from every data train. Once the system surpassed 32 
{\it DataTrain}s this became a limiting factor.

\begin{figure*}
\begin{center}
\includegraphics[scale=0.4]{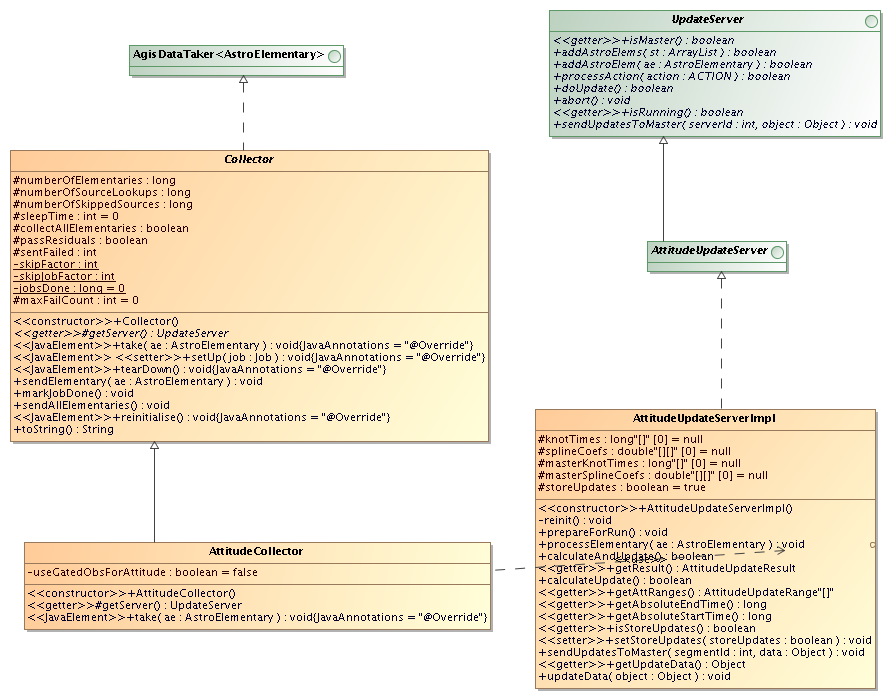}
\end{center}
\caption{\label{fig:attUpd} UML diagram for the distributed attitude update. 
The calibration update has similar classes.}
\end{figure*} 

On each data train an {\it AttitudeCollector} is registered. This gathers all
of the elementaries and passes them to the appropriate 
{\it AttitudeUpdateServer}. Appropriate here means the attitude server dealing
with the time bin in which the observation falls. In some cases the segments
overlap and an observation must be sent to two servers simultaneously. 
Again RMI is used for this passing and the observations carry the updated
source parameters with them. 

The {\it AttitudeUpdateServer}(s) adds to the partial equations for each
observation passed. It must wait until the end of the run to ensure
all observations have been seen before doing the final computation. The end of
the run is signalled, via RMI, by the {\it RunManager}. At this point the updated
spline coefficients are calculated and written to the {\it Store}.
The server also now sends the updated attitude to the {\it ConvergenceMonitor} 
so it may be plotted on the website.

\subsection {Calibration update}\label{sect:calupd}

The geometric calibration model  deals with the precise placement and
orientation of the CCDs in the focal plane. Within the optical system light bounces
off six highly polished mirrors before hitting the CCDs in the Focal Plane Assembly.
Since there are no on-board calibration devices 
 a distortion in a mirror  is indistinguishable from a displacement of a
CCD. In both cases the image centroid will not appear where it should be. This also means
that any such shift can be modelled in terms of CCD orientation, ignoring the mirrors 
entirely, and this is precisely what we do in AGIS.

Geometric calibration parameters for the
CCDs, such as orientation, scale and mechanical distortions, are defined
on timescales of hours or months as needed and are know as {\em CalibrationEffect}s.
This transformation for Gaia is quite involved \citet[see][]{2005A&A...438..745B}, 
yet for our purposes we may 
consider an instantaneous position $\eta^\text{obs}$ for the source in the field of view.
We define the astrometric calibrations in the following generalised form:
\begin{equation}
\eta^\text{obs}_l =
\eta_n^0 +\sum_r E_r(l)\label{eq:etaobs} \, ,
\end{equation}
where $l$ is the observation index and each of the $E_r(l)$ represents one basic 
{\em CalibrationEffect}, being a linear combination of calibration functions
$\Phi_{rs}(l)$: 
\begin{equation}
E_r(l) = \sum_s c_{rs} \Phi_{rs}(l)\label{eq:ei} \, .
\end{equation}
The coefficients $c_{rs}$ constitute the whole set of calibration parameters. 
In the calibration update we solve these coefficients by a least-squares 
system similar to Eq.~(\ref{eq:lsobsmatrix}).

The functions $\Phi_{rs}$ receive the observation index $l$
and it is assumed that this index suffices to derive whatever dependencies 
are needed to evaluate the corresponding function/effect for this
observation.  
Examples of such dependencies are: the telescope index (preceding/following
field of view); CCD row number; CCD strip number; pixel column within the CCD;
time; relevant astrometric, photometric, and spectroscopic source parameters;
auxiliary parameters (e.g., optical background level, illumination history of the pixel
column). In this generic calibration scheme the dependencies are not hardcoded, 
and we do not know exactly how many calibration parameters there will be in the 
mission. 
Furthermore the calibration effects are all specified in an XML file allowing 
for easy addition (or removal) of specific effects in an AGIS execution. 

\begin{figure*}
\begin{center}
\includegraphics[scale=1.5]{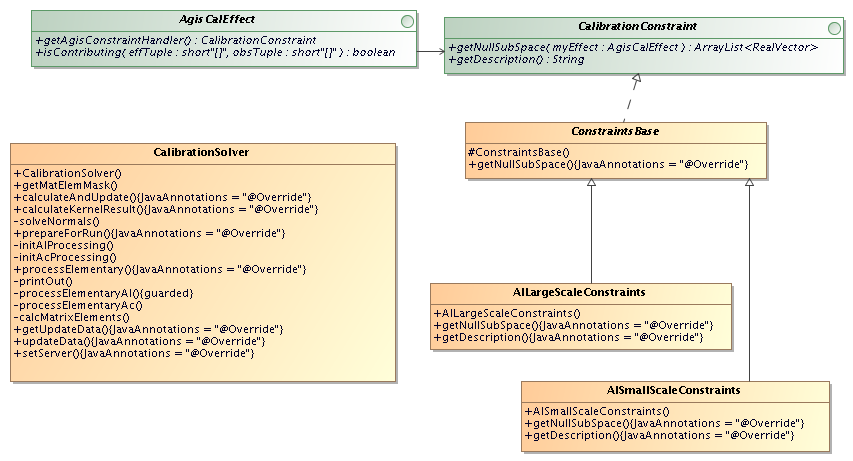}
\end{center}
\caption{\label{fig:calunit} UML for the {\em CalibrationEffect}s}
\end{figure*}

Following the terminology introduced in \citet{LL08}, the calibration parameters 
can be grouped into {\em calibration units} that can be handled separately because
any given observation $l$ can only belong to one calibration unit.
Within a calibration unit, on the other hand, each observation typically contribute
to many different effects, for example to irregularities both on a large scale (e.g., 
between CCDs) and on a small scale (e.g., between pixel columns). 
Our estimate is that no calibration unit will have more than about 10,000 parameters, 
which is negligible compared to the attitude parameters.
Still, the memory requirements in the calibration block are larger than in the attitude
update because there is no obvious way to exploit the sparseness of the
normal matrix within each calibration unit.
The {\rm CalibrationEffect}s are depicted using UML in Fig.~\ref{fig:calunit}.

From the perspective of distributed processing one must consider that, 
unlike attitude, here an observation will end up going to many calibration effects,
e.g., both the large-scale and the small-scale calibration.
We may however process all effects for a row of CCDs on a separate machine. 
The processing for calibration is not a huge overhead; as for attitude, the main
bottleneck is the sending of all observations to the calibration server(s). Unlike
the attitude, the calibration server can process the incoming observations more
quickly and it has not been an overall bottleneck in the system.

The framework is similar to that of the attitude update. A {\it CalibrationCollector} is
registered with each {\it DataTrain}, collects the required observation
information and sends it to the {\it CalibrationUpdateServer} via RMI. 
The server accumulates the equations during the run and performs the final
calculation when signalled by the {\it RunManager} that the run is complete. 
It writes the updated calibrations to the {\it Store} and sends them to the
{\it GisConvergenceMonitor} for plotting on the website.

\subsection{Global update}

The global parameters, nominally some of the Parameterized Post-Newtonian (PPN) 
relativistic parameters, are estimated using a robust least-squares algorithm similar
to Eq.~(\ref{eq:lsobsmatrix}) but now involving all the observations but only a (very)
small set of parameters. The treatment is practically identical to a calibration parameter 
which spans the entire mission. As such it would be possible to combine this with the
calibration update in a later version of the system, but for other reasons it is convenient 
to separate these terms, for example, to more easily estimate their correlations. 
As in the case for the attitude and calibration blocks we also have a {\it GlobalCollector} 
and a {\it GlobalUpdateManager} functioning in the same manner as described
previously.

We will have sufficient observations and full sky 
coverage to decouple the global parameters from the astrometric parameters. 
Currently we only calculate PPN-$\gamma$ due to solar system body deflection, but other variants 
will be added in the future, for example, separate and combined values of PPN-$\gamma$ 
due to deflection by the major planets. The calculation of additional global parameters can provide a 
sanity check on the entire solution, i.e., a value wildly departing from the nominal value in 
the simulation data can only mean we are doing something very wrong somewhere.

\subsection{Secondary source update}\label{sec:secondary}

Nominally the entire data set could be put through AGIS, however we know that many 
binary stars and other complex objects 
will not work well with the simple observation model used. Hence 
only a fraction (between 10 and 50\%) of the sources observed by Gaia are processed in 
AGIS. The selection of these {\it primary sources} will be done partly based on information
from other parts of the processing chain (e.g., detected double stars), but mainly from
the goodness-of-fit statistics gathered while performing a trial source update. If the
fit is bad for the source, it is not accepted as a primary source but relegated to
secondary source status. The selection of primary/secondary sources is itself
an iterative process, which must be repeated after more accurate estimates have
been obtained of the attitude and calibration parameters.
 
The AGIS solution, thus based on a `clean' subset of the sources, provides an accurate 
celestial reference frame along with a correspondingly accurate attitude and 
geometric calibration. These outputs will be used to update the remaining fraction
of the sources. This secondary star update is effectively identical to the
source update block described in Sect.~\ref{sect:sourceupd} but must only be run once 
over the data. This secondary solution will still
not make sense for all types of objects (e.g., resolved binaries), which will 
be picked up in other parts of the processing chain.

\section{Results} \label{sect:results}

Some run times for the system are given in Table~\ref{tab:agistimes}. AGIS
has been running almost continuously since the end of 2005 on different simulated
data sets. 
The current system requires around 40 iterations to remove initial (random and systematic)
catalogue errors of about 100~milliarcseconds, based on the simulated observations. 
This level of initial errors is well above expected mission levels. After 40 iterations 
AGIS the source errors have been reduced to a level that is consistent with the observational 
noise level, i.e., some microarcsec for the brighter sources. Moreover, none of the systematic
errors introduced in the starting values remain in the converged solution. A more comprehensive
study of the results from AGIS will be provided in another paper.

\section{Conclusion}
The overall AGIS architecture and many of the components have been described in
some detail. This is a software system designed and optimised to perform the
Gaia astrometric data reduction involving the solution of a system with hundreds
of millions of parameters and hundreds of billions of observations.

Advanced features of the Java language have been
employed to make this system work well and remain very portable. Despite
skepticism we have found Java reliable and robust, and sufficiently 
performant for our purposes. 
More work is needed in the coming years to further optimise AGIS, 
but a very good system is already in place and well understood. 

\begin{acknowledgements}

The constant work of the Gaia Data Processing and Analysis Consortium (DPAC) has
played an important part in this work. We are particularly indebted to CU2 for
the production of independently simulated Gaia-like data for use in the system.
The data simulations have been done in the supercomputer Mare Nostrum at Barcelona
Supercomputing Center -- Centro Nacional de Supercomputaci{\'o}n
(The Spanish National Supercomputing Center).
Research and development in Sweden is kindly supported by the Swedish National
Space Board (SNSB).

Thanks to  Xavier Luri for his guidance and input on early versions of the text.  

Our special thanks go to Gaia's former Project Scientist
Michael Perryman, whose vision, leadership, and enthusiasm in the early years
of the project laid the foundations for the excellent progress that is today seen
throughout DPAC and with AGIS in particular.

\end{acknowledgements}

\bibliographystyle{spbasic} 
\bibliography{agis} 

\end{document}